\documentstyle[epsfig,11pt,newpasp,twoside]{article}
\index{pulsars}
\index{gamma-rays}
\index{GLAST}

\def\edcomment#1{\iffalse\marginpar{\raggedright\sl#1\/}\else\relax\fi}
\reversemarginpar

\begin{document}
\title{Future Facilities for Gamma-Ray Pulsar Studies}
 \author{D. J. Thompson}
\affil{NASA Goddard Space Flight Center, Greenbelt, Maryland, USA, and GLAST Large Area Telescope Collaboration}

\begin{abstract}
Pulsars seen at gamma-ray energies offer insight into particle acceleration to very high energies, along with information about the geometry and interaction processes in the magnetospheres of these rotating neutron stars.  During the next decade, a number of new gamma-ray facilities will become available for pulsar studies.  This brief review describes the motivation for gamma-ray pulsar studies, the opportunities for such studies, and some specific discussion of the capabilities of the Gamma-ray Large Area Space Telescope (GLAST) Large Area Telescope (LAT) for pulsar measurements. 
\end{abstract}

\section{Introduction}

Pulsars represent astrophysical laboratories for extreme conditions. The high densities, temperatures, velocities, electric potentials, and magnetic fields associated with these spinning neutron stars give rise to high-energy emission through a variety of mechanisms.  In particular, gamma-rays are produced by interactions of particles accelerated to the highest energies by electromagnetic and shock processes.  Gamma-ray pulsar data therefore complement observations from longer wavelengths, where the emission often originates from thermal or plasma processes. 

\section{Motivation - Current Knowledge of Gamma-Ray Pulsars}

This discussion of future gamma-ray facilities is motivated strongly by the knowledge gained from previous missions, most notably the instruments on the Compton Gamma Ray Observatory (CGRO) - BATSE, OSSE, COMPTEL, and EGRET.  Combined with what has been learned at radio, optical, and X-ray wavelengths, the gamma-ray data help illustrate the multiwavelength nature of at least some pulsars. Figure 1 is an example, showing the light curves of the seven pulsars with the strongest gamma-ray signals.

\begin{figure}[b!] 
\centerline{\epsfig{file=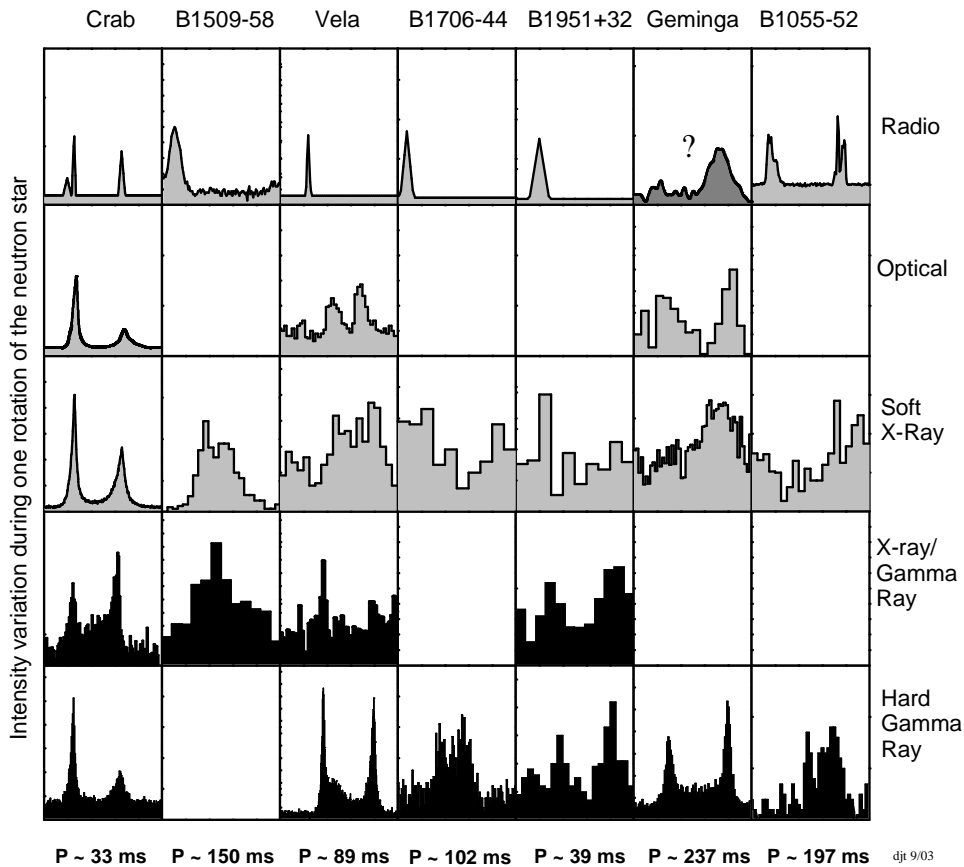,width=5.1in}}
\vspace{10pt}
\caption{Light curves of seven gamma-ray pulsars in five energy bands,  from left to right in order of characteristic age. Each panel shows one full rotation of the neutron star. The X-ray and gamma-ray references are those from Fig. 1 of Thompson (2001), with the addition of the X-ray data for PSR B1706$-$44 from Gotthelf, Halpern, and Dodson (2002).  Radio references (left to right): Manchester, 1971; Ulmer et al., 1993; Kanbach et al., 1994; Johnston et al., 1992; Kulkarni et al., 1988, Kuz'min and Losovskii,  1997; Fierro et al., 1993. Optical references: Crab, Groth, 1975; Vela, Manchester et al., 1980; Geminga, Shearer et al., 1998.}
\label{fig1}
\end{figure}
 
What seems clear from this figure is that pulsar light curves show at least as much variability from energy band to energy band for a given pulsar as from pulsar to pulsar within a single energy band.  The combination of physics and geometry is strongly wavelength-dependent, but not in any simple way.  Perhaps the most consistency is seen in the high-energy gamma-ray band, where all six of the pulsars that are seen have double-pulsed light curves.  It can be argued that this predominance of double pulses is more consistent with emission from a hollow cone or other surface above a single magnetic pole than from emission coming from both magnetic poles.

In addition to these seven high-confidence gamma-ray pulsars, at least three other radio pulsars show some evidence of pulsed gamma-ray emission, and a number of other gamma-ray sources are pulsar candidates based on energetics, position, and in some cases spectral information.  Beyond the pulsars themselves, gamma-ray emission from supernova remnants provides further information about particle acceleration in pulsar environments. 

\begin{figure}[b!] 
\centerline{\epsfig{file=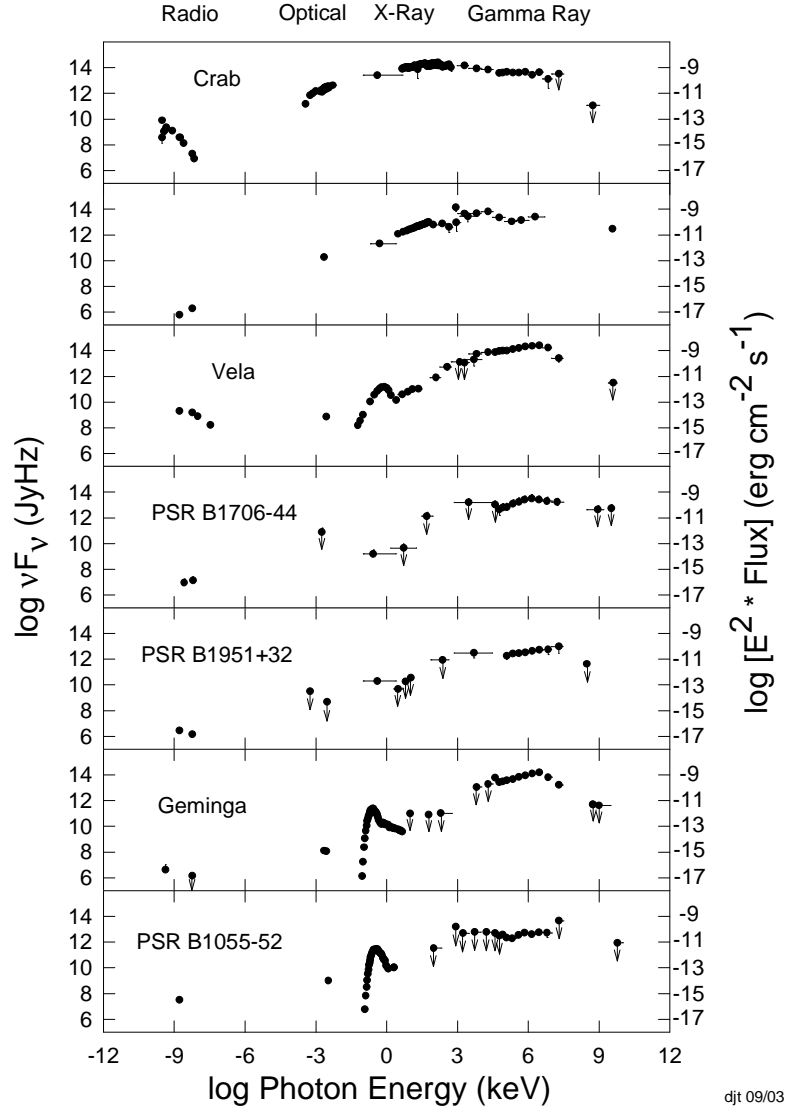,width=4.8 in}}
\vspace{10pt}
\caption{Multiwavelength spectra of seven gamma-ray pulsars.  Updated from Thompson et al. (1999) with data from Fierro et al. (1998) for the Crab (gamma rays) and Jackson et al. 2002 for Geminga (hard X-rays). }
\label{fig2}
\end{figure}
Figure 2 summarizes the broad-band energy spectra for these seven pulsars. The data, shown in $\nu$F$\nu$ format (or E$^2$ times the photon number spectrum), indicate the observed power per frequency interval across the spectrum.  Some features of these spectra are:

\begin{itemize}
\item 
  In all cases, the maximum observed power output is the in the gamma-ray band. 

\item Vela, Geminga, and B1055$-$52 all show evidence of a thermal component in X-rays, thought to be from the hot neutron star surface. 

\item Each of these pulsars exhibits a high-energy spectral turnover or spectral break. No pulsed emission is seen above 30 GeV, the upper limit of the Compton Observatory/EGRET observations. These changes in spectral index appear to be related to the calculated surface magnetic field of the pulsar.  The nature of the high-energy cutoff is an important feature of pulsar models. 

\end{itemize}

Compared to the vast number of known radio pulsars, the statistics for gamma-ray pulsars are meager.  Nevertheless, these observations provide ample motivation for future gamma-ray missions to study pulsars and their surroundings.  For additional information about gamma-ray pulsar observations, see Thompson (2003), and for a summary of the relationship between gamma-ray and radio pulsars, see Kramer et al. (2003)

\section{Present and Future Gamma-Ray Opportunities}

\subsection{INTEGRAL}

The INTErnational Gamma-Ray Astrophysics Laboratory (INTEGRAL) is a European-led mission that had its launch on October 17, 2002, into an elliptical 72 hour orbit.  For a summary of the INTEGRAL mission, see Winkler et al. (2003) or http://astro.estec.esa.nl/Integral/isoc/.  Some relevant aspects of the project are:

\begin{itemize}
\item 
  Two gamma-ray telescopes, an imager and a spectrometer, cover the energy range 15 keV to 10 MeV. 

\item Good timing has enabled pulsar studies.  

\item High-resolution gamma-ray line spectrometry allows searches for radioactive element decay, such as Ti-44, in supernova remnants.

\item INTEGRAL also carries an optical telescope and an X-ray telescope

\item The mission lifetime is two years minimum, with plans for up to five years. 

\end{itemize}

\subsection{Swift}

Although the international Swift mission is primarily focused on studies of gamma-ray bursts, it will have gamma-ray capabilities useful for pulsar studies as well.  Current information about Swift, currently planned for launch in May 2004, can be found at http://swift.gsfc.nasa.gov/.  Some important aspects of Swift are:

\begin{itemize}
\item 
  A wide-field gamma-ray telescope covers the energy range 15 - 150 keV. 

\item Good timing capabilities will be useful for pulsar studies, especially magnetars, which are thought to be related to soft gamma repeaters.  

\item Swift also carries an optical/UV telescope and an X-ray telescope

\item The mission lifetime is two years minimum, with options to extend. 

\end{itemize}

\subsection{AGILE}

Astro-rivelatore Gamma a Immagini LEggero (AGILE) is a small gamma-ray telescope being built by an Italian collaboration for launch in mid-2005.  Current information about AGILE can be found at http://agile.mi.iasf.cnr.it/Homepage/index.shtml.  Some important aspects of AGILE are:

\begin{itemize}
\item 
  This wide-field high-energy gamma-ray telescope covers the energy range 20 MeV - 50 GeV, similar to that of CGRO/EGRET and GLAST/LAT. AGILE will have somewhat better sensitivity than EGRET.

\item Accurate timing will facilitate pulsar studies.  

\item Super-AGILE is a co-aligned hard X-ray telescope covering the energy range 10 - 40 keV. 

\item The mission lifetime is two years minimum, with options to extend. 

\end{itemize}

\subsection{Ground-Based Gamma-Ray Telescopes}

Complementary to the satellite gamma-ray telescopes just described are the new ground-based gamma-ray telescopes that operate at higher energies than can typically be reached from space.  Although none of the previous generation of such telescopes detected any pulsed gamma radiation, there is evidence for very-high-energy emission from supernova remnants.  The new telescopes, including MILAGRO, HESS, STACEE, CANGAROO-3, MAGIC, and VERITAS, give greater sensitivity and in many cases a lower energy threshold that will overlap that of GLAST.  In contrast to the satellite telescopes, which are often photon-limited, the ground-based telescopes have high statistics but are background limited.  For additional information about these, see Rowell (2003).

\subsection{GLAST}

The Gamma-ray Large Area Space Telescope (GLAST) is a major international gamma-ray mission led by NASA and the Department of Energy in the U.S. and including substantial contributions from many other countries. Its two wide-field instruments cover the energy range from about 10 keV to 300 GeV. Scheduled for launch in late 2006 or early 2007, GLAST is planned for a five year mission or longer. Information about the project can be found at http://glast.gsfc.nasa.gov/. 

In terms of pulsars and their environments, the principal GLAST contribution will come from the high-energy Large Area Telescope (LAT), which was originally called GLAST by itself.  The other instrument, the GLAST Burst Monitor (GBM), does not have sufficient photon timing or resolution to add substantially to pulsar studies.  The following discussion will concentrate, therefore, on the properties of the LAT (Gehrels and Michelson (1999), and http://www-glast.stanford.edu/):

\begin{itemize}
\item 
  The LAT covers the energy range 20 MeV - 300 GeV, with sensitivity at least a factor of 30 greater than EGRET.

\item The LAT energy reach brings good sensitivity to the largely-unexplored range above 10 GeV, where EGRET had limited capability. 

 \item The huge LAT field of view ($>$ 2 ster) and planned operation in a scanning mode will provide exposure to most of the sky every day. 

\item The point spread function will be substantially better than EGRET's, and the capability to localize sources will be better than 1 arcmin for many sources. 

\item Accurate pulsar timing is part of the baseline design.  

\item The LAT has no expendable components that could limit the life of the instrument.

\end{itemize}

Figure 3 shows one measure of pulsar observability: the pulsar spin-down luminosity divided by 4$\pi$ times the square of the 
distance, the total available pulsar energy output at Earth.  This plot uses the new ATNF pulsar catalogue (http://www.atnf.csiro.au/research/pulsar/psrcat/) with distance estimates using the new NE2001 electron density model of Cordes and Lazio (2002).
 Ten gamma-ray pulsars and candidates are shown as the large squares.  Six of the seven pulsars with the 
highest value of this observability parameter are gamma-ray pulsars. Below these, only a handful of 
pulsars are visible in gamma rays.  The GLAST LAT sensitivity will push the lower limit 
down substantially farther.  Two approximate sensitivity limits are shown for GLAST LAT - one for 
low-Galactic-latitude sources and one for those at high latitudes, because the high 
diffuse emission along the Galactic plane reduces the sensitivity for point source 
detection.  The phase space that GLAST LAT opens up is substantial, particularly when coupled with the new radio pulsar discoveries that have been made and will be made in the near future. 

\begin{figure}[b!] 
\centerline{\epsfig{file=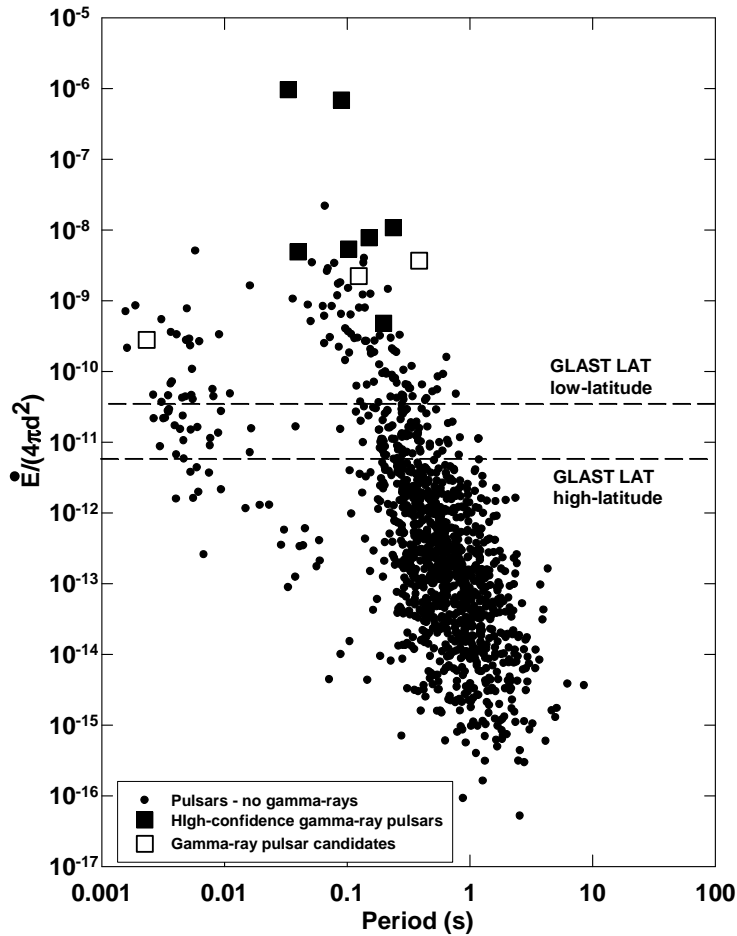,width=4in}}
\vspace{10pt}
\caption{Gamma-ray pulsar observability, as measured by the spin-down energy seen at earth.}
\label{fig3}
\end{figure}

A second important feature of the LAT is that its sensitivity will be large enough to allow pulsar searches using the gamma-ray data alone. Various analyses have shown that only the brightest pulsars could be found in 
the EGRET data using blind searches (Brazier and Kanbach 1996; Jones, 1998; Chandler et al. 2001). For all of the other 
unidentified EGRET sources, the photons are just too few and too far apart in time to derive 
unambiguous pulse periods.  With GLAST LAT, periodicity searches will be feasible for 
most, if not all, the unidentified EGRET sources (Mattox, 2000; McLaughlin and Cordes, 2000).  The potential is to find a whole 
new population of rotation-powered pulsars, much as X-ray astronomy has started to 
do in the past few years.  Radio-quiet gamma-ray pulsars could be an important new window onto the physics of 
the extreme conditions around these spinning neutron stars.

Despite the search capability of the LAT, the most powerful test of pulsar detection will remain the analysis of the gamma-ray data with known timing solutions.  The LAT, like EGRET, will be constrained by statistics, with times between photons representing many thousand rotations of the neutron stars in some cases.  In order to obtain the maximum sensitivity, therefore, GLAST will need contemporaneous timing solutions.  A program to coordinate radio observations during the GLAST operating epoch is being organized by Steve Thorsett, one of the GLAST InterDisciplinary Scientists (IDS).

The number of pulsars that will be seen by GLAST LAT or other new gamma-ray telescopes is highly uncertain, as are predictions of which known radio pulsars should be visible at gamma-ray energies.  The reason for such uncertainty is that a variety of models for gamma-ray pulsars have developed, and these models differ on their expectations for future observations.  Clearly one important goal of these gamma-ray observations will be to challenge and refine the models. 

GLAST is planned as a facility mission, with data available to all scientists after an initial sky survey, along with software to perform standard analyses, including pulsar timing.  A GLAST Science Support Center is located at Goddard Space Flight Center to help with data analysis.

\section{Summary}

Gamma rays have become part of the multiwavelength approach to the study of pulsars. The relatively large power output seen in gamma rays makes them especially useful probes of the particle acceleration and interaction processes in pulsar magnetospheres.  New space-based and ground-based gamma-ray missions are emerging that will add substantially to the capabilities in this high energy range.

\end{document}